\documentclass[conference]{IEEEtran}
\usepackage{graphicx}
\usepackage{cite}
\usepackage{amsmath,amssymb,amsfonts}
\usepackage{algorithmic}
\usepackage{textcomp}
\usepackage{xcolor}
\usepackage{svg}
\usepackage{xspace}
\usepackage[acronym]{glossaries}
\makeglossaries
\usepackage{multirow}
\usepackage{booktabs} 
\usepackage{url}
\usepackage{color}
\usepackage{multirow}
\usepackage{enumitem}
\usepackage{array}
\usepackage[nolist,nohyperlinks]{acronym}
\usepackage{footnote}
\makesavenoteenv{tabular}

\usepackage{fancyhdr}
\def\BibTeX{{\rm B\kern-.05em{\sc i\kern-.025em b}\kern-.08em
    T\kern-.1667em\lower.7ex\hbox{E}\kern-.125emX}}

\acrodef{mIoU}{mean Intersection over Union}
\acrodef{DL}{deep learning}
\acrodef{ML}{machine learning}
\acrodef{DSC}{dice coefficient}
\acrodef{CRC}{Colorectal cancer}

\begin{document}
\title{TransResU-Net: Transformer based ResU-Net for Real-Time Colonoscopy Polyp Segmentation}
\author{\IEEEauthorblockN{Nikhil Kumar Tomar \IEEEauthorrefmark{1}, 
Annie Shergill, M.D.\IEEEauthorrefmark{2},
Brandon Rieders M.D.\IEEEauthorrefmark{3},
Ulas Bagci, Ph.D\IEEEauthorrefmark{4},
Debesh Jha, Ph.D\IEEEauthorrefmark{4}}

\IEEEauthorblockA{\IEEEauthorrefmark{1} School of Computer and Information Sciences, Indira Gandhi National Open University\\ 
\IEEEauthorrefmark{2}Larkin Community Hospital Palm Springs Campus, USA\\
\IEEEauthorrefmark{3} Yea Long Island Jewish Valley Stream, USA\\
\IEEEauthorrefmark{4} Department of Radiology, Feinberg School of Medicine, Northwestern University, USA\\ 
}
}

\maketitle
\thispagestyle{fancy}
\begin{abstract}
Colorectal cancer (CRC) is one of the most common causes of cancer and cancer-related mortality worldwide. Performing colon cancer screening in a timely fashion is the key to early detection. Colonoscopy is the primary modality used to diagnose colon cancer.  However, the miss rate of polyps, adenomas and advanced adenomas remains significantly high. Early detection of polyps at the precancerous stage can help reduce the mortality rate and the economic burden associated with colorectal cancer. Deep learning-based computer-aided diagnosis (CADx) system may help gastroenterologists to identify polyps that may otherwise be missed, thereby improving the polyp detection rate. Additionally, CADx system could prove to be a cost-effective system that improves long-term colorectal cancer prevention. In this study, we proposed a deep learning-based architecture for automatic polyp segmentation, called Transformer ResU-Net (TransResU-Net). Our proposed architecture is built upon residual blocks with ResNet-50 as the backbone and takes the advantage of transformer self-attention mechanism as well as dilated convolution(s). Our experimental results on two publicly available polyp segmentation benchmark datasets showed that TransResU-Net obtained a highly promising dice score and a real-time speed. With high efficacy in our performance metrics, we concluded that TransResU-Net could be a strong benchmark for building a real-time polyp detection system for the early diagnosis, treatment, and prevention of colorectal cancer. The source code of the proposed TransResU-Net is publicly available at \url{ https://github.com/nikhilroxtomar/TransResUNet}.


\end{abstract}


\section{Introduction}
\ac{CRC} is the third most common cancer affecting men and the second most common cancer affecting women globally according to the World Health Organization GLOBOCAN database~\cite{sung2021global}. Approximately 70\% of cases occur in the colon and the remaining occur in the rectum~\cite{siegel2022cancer}. Considering that colonoscopy is the primary screening modality and most prevalent diagnostic technique in gastrointestinal endoscopy,  quality assurance is critical~\cite{seeff2004many}. Detection of cancer at an early and curable stage and removal of precancerous adenomas or serrated lesions during colonoscopy is the key to colon cancer diagnosis. It is also associated with reduction in mortality~\cite{kronborg1996randomised,zauber2012colonoscopic}. 

Colonoscopy is an expensive, resource-demanding, and unpleasant procedure. Many patients show unwillingness to participate in \ac{CRC} screening program repeatably. According to a recent meta-analysis, up to 26\% of colonoscopies may have missed lesions and adenomas~\cite{zhao2019magnitude}. This is because it is an operator driven procedure and solely dependent on the clinical acumen and skills of the endoscopist. With the current colonoscopy equipment, less experienced endoscopists cannot distinguish all of neoplastic and non-neoplastic polyps during routine colonoscopy examination~\cite{wadhwa2020physician,dayyeh2015asge}. An automatic algorithm based real-time diagnosis of polyps (irrespective of their morphology) during colonoscopy could help endoscopists in identifying potential polyps for removal with improved efficiency and accuracy. It also reduces the access barrier to pathological services~\cite{wilson2018access}.  

\begin{figure*}[t!]
    \centering
    \includegraphics[width=0.9\textwidth]{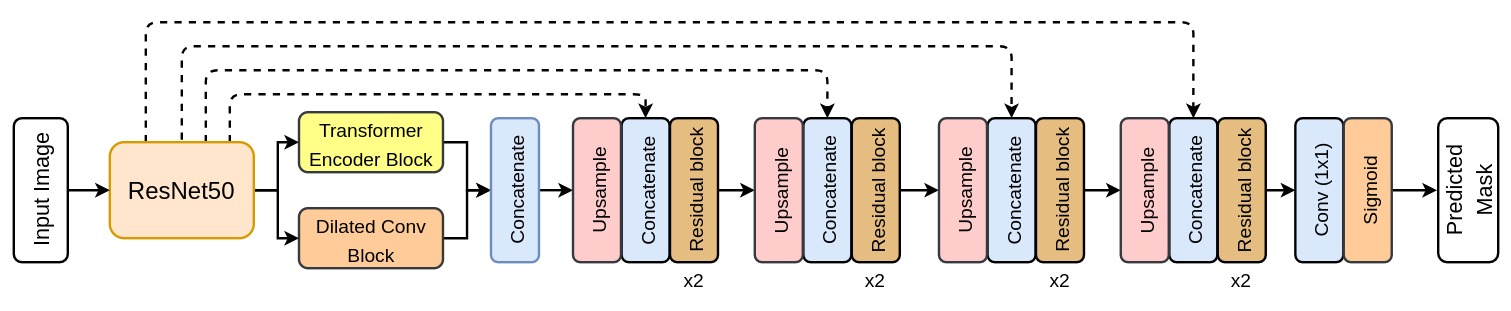}
    \caption{Block diagram of the proposed TransResU-Net architecture}
    \label{fig:proposed-architecture}
\end{figure*}

Recently, there has been a great interest in deep learning in \ac{CRC} screening. Various studies aimed to develop CADx models for automatic polyp segmentation~\cite{tomar2022tganet,huang2021hardnet,jha2020doubleu,fan2020pranet,tomar2022fanet,jha2021real,zhou2018unet++}. Among the recent deep learning architecture, Transformer~\cite{vaswani2017attention} and U-Net~\cite{ronneberger2015u} based architecture have attracted the most attention. Various extension of UNet have been proposed in the literature~\cite{zhou2018unet++,jha2019resunet++,jha2020doubleu} for automatic polyp segmentation. Despite good results produced by these studies, more research needs to be done on descent sized polyp datasets to demonstrate the effectiveness of the proposed method for automatic polyp segmentation. One of the promising extensions of UNet is ResUNet~\cite{zhang2018road}. The architecture is built on residual units that uses identity mapping (shortcut connection). The residual unit eases the training of the deep neural network whereas the identity mapping facilitate the better flow of the gradients. Similarly, Yu et al.~\cite{yu2015multi} presented an efficient dilation convolution block to increase the context module, which helps to improve the accuracy of the semantic segmentation network. 

\begin{table*}[t!]
\centering
\caption{Quantitative results on the on Kvasir-SEG~\cite{jha2020kvasir}.}
 \begin{tabular} {@{}l|c|c|c|c|c|c|c|c@{}}
\toprule
\textbf{Method} & \textbf{Backone} &\textbf{DSC} & \textbf{mIoU} &\textbf{Recall}  &\textbf{Precision}& \textbf{Accuracy} &\textbf{F2} &\textbf{FPS}\\ 
\hline
U-Net~\cite{ronneberger2015u} &-	&0.8264	&0.7472	&0.8504	&0.8703	&0.9510	&0.8353	&156.83\\
ResU-Net~\cite{zhang2018road}&-&	0.7642	&0.6634	&0.8025	&0.8200	&0.9341	&0.7740	&\textbf{196.85}\\
U-Net++~\cite{zhou2018unet++}&-&0.8228	&0.7419	&0.8437	&0.8607	&0.9491	&0.8295	&126.14\\
ResU-Net++~\cite{jha2019resunet++}&- &0.6453	&0.5341	&0.6964	&0.7080	&0.9044	&0.6575	&57.99\\
ColonSegNet~\cite{jha2021real}	&-&0.7920	&0.6980	&0.8193	&0.8432	&0.9415	&0.7999	&129.04\\
HarDNet-MSEG~\cite{huang2021hardnet}&-&0.8260 &0.7459	&0.8485	&0.8652	&0.9492	&0.8358	&42.00\\
DeepLabV3+\cite{chen2018encoder}  &ResNet50 &0.8837	&0.8173	&0.9014	&\textbf{0.9028}	&\textbf{0.9679}	&0.8904	&102.62\\
DDANet~\cite{tomar2021ddanet}	&- &0.7415	&0.6448	&0.7953	&0.7670	&0.9326	&0.7640	&88.70\\
\textbf{TransResU-Net (Ours)}	& ResNet50 &\textbf{0.8884} &\textbf{0.8214} &\textbf{0.9106} &0.9022 &0.9651 &\textbf{0.8971} &48.61\\
\hline
\end{tabular}
\vspace{-2mm}
\label{tab:results-kvasir}
\end{table*}

Inspired by the successes of Transformers~\cite{demir2022transformer}, residual unit~\cite{he2016deep},  and dilated convolution~\cite{yu2015multi}, we develop a novel deep learning-based architecture, TransResU-Net. We tested the performance of TransResU-Net on two decent sized publicly available polyp datasets. It is to confirm if the proposed method can detect early signs of \ac{CRC} with high performance and a real-time speed. The main contribution of our work can be summarized as follows: 
\begin{enumerate}
    \item We have proposed a novel deep segmentation architecture called TransResU-Net, which combines the strengths of the  transformer block, dilated convolution layers with the pre-trained ResNet50, which has never been done before. 
    
    \item We compared TransResU-Net with eight commonly used benchmark algorithms for the automated polyp segmentation tasks. TransResU-Net showed state-of-the-art performance on Kvasir-SEG~\cite{jha2020kvasir} and BKAI-IGH~\cite{lan2021neounet} dataset. 
\end{enumerate}

\section{Method}
\vspace{-1mm}
Figure~\ref{fig:proposed-architecture} show the block diagram of our proposed TransResU-Net. The proposed architecture follows an encoder-decoder scheme, where we have a pre-trained ResNet50 as the encoder and four decoder blocks. The input image is fed to the pre-trained encoder, which consists of multiple bottlenecks residual blocks along with pooling layers which transform the input image into a spatially reduced feature representation. The output from the pre-trained encoder is then passed through a transformer encoder block and a dilated convolution block. The transformer encoder block~\cite{vaswani2017attention} consists of a self-attention network which is followed by a feed-forward neural network which helps the proposed TransResU-Net to learn a more robust representation. Meanwhile the dilated convolution block helps the convolution filters to increase their receptive field and thus enhance the effective capacity of the network.

The dilated convolution block consists of four parallel $3\times3$ convolution layers, where each layer has a dilation rate of $1$, $3$, $6$, and $9$ respectively.  These layers are then followed by batch normalization and a ReLU activation function. Next, we concatenate the features from all four layers and pass them through a $1\times1$ convolution layer to effectively reduce the number of feature channels. The output from both the transformer encoder block and the dilated convolution block are concatenated and passed to the first decoder block. The decoder block begins with a bilinear upsampling, which is followed by the concatenation with the skip connection from the encoder block. These skip connections help to get the feature maps directly from the encoder to the decoder block, which is important since some of the features are lost due to the depth of the network. These skip connections also help in better flow of the gradients during the backpropagation and thus help to improve the overall performance of the network. These concatenated feature maps are then passed through two residual blocks, which consist of the two $3\times3$ convolution layers and an identity mapping. Subsequently, the output from the first decoder block is passed to the second decoder block and so on. This way the feature maps are progressively transformed to more meaningful semantic features. The output from the last decoder block is passed through a $1\times1$ convolution layer followed by a sigmoid activation function which generates a binary segmentation mask.

\begin{table*}[t!]
\centering
\caption{Quantitative results on the on BKAI-IGH~\cite{lan2021neounet}.}
 \begin{tabular} {@{}l|c|c|c|c|c|c|c|c@{}}
\toprule
\textbf{Method} & \textbf{Backone} &\textbf{DSC} & \textbf{mIoU} &\textbf{Recall}  &\textbf{Precision}& \textbf{Accuracy} &\textbf{F2} &\textbf{FPS}\\ 
\hline
U-Net~\cite{ronneberger2015u} &-	&0.8286	&0.7599	&0.8295	&0.8999	&0.9903	&0.8264	&160.27\\
ResU-Net~\cite{zhang2018road}&-	&0.7433	&0.6580	&0.7447	&0.8711	&0.9843	&0.7387	&\textbf{197.94}\\
U-Net++~\cite{zhou2018unet++}&-&0.8275	&0.7563	&0.8388	&0.8942	&0.9895	&0.8308	&123.45\\
ResU-Net++~\cite{jha2019resunet++}&- &0.7130	&0.6280	&0.7240	&0.8578	&0.9832	&0.7132	&55.86\\
ColonSegNet~\cite{jha2021real}	&-&0.7748	&0.6881	&0.7852	&0.8711	&0.9843	&0.7746	&122.42\\
HarDNet-MSEG~\cite{huang2021hardnet}&-&0.7627	&0.6734	&0.7532	&0.8344	&0.9863	&0.7528	&41.20\\
DeepLabV3+\cite{chen2018encoder}&ResNet50 &0.8937	&0.8314	&0.8870	&0.9333	&0.9937	&0.8882	&99.16\\
DDANet~\cite{tomar2021ddanet}	&- &0.7269	&0.6507	&0.7454	&0.7575	&0.9851	&0.7335	&86.46\\
\textbf{TransResU-Net (Ours)}	& ResNet50  &\textbf{0.9154} &\textbf{0.8568} &\textbf{0.9142} &\textbf{0.9299} &\textbf{0.9938} &\textbf{0.9129} & 42.09 \\
\hline
\end{tabular}
\label{tab:results-bkai}
\end{table*}

\begin{figure*}[t!]
    \centering
    \includegraphics[width = 0.55\textwidth]{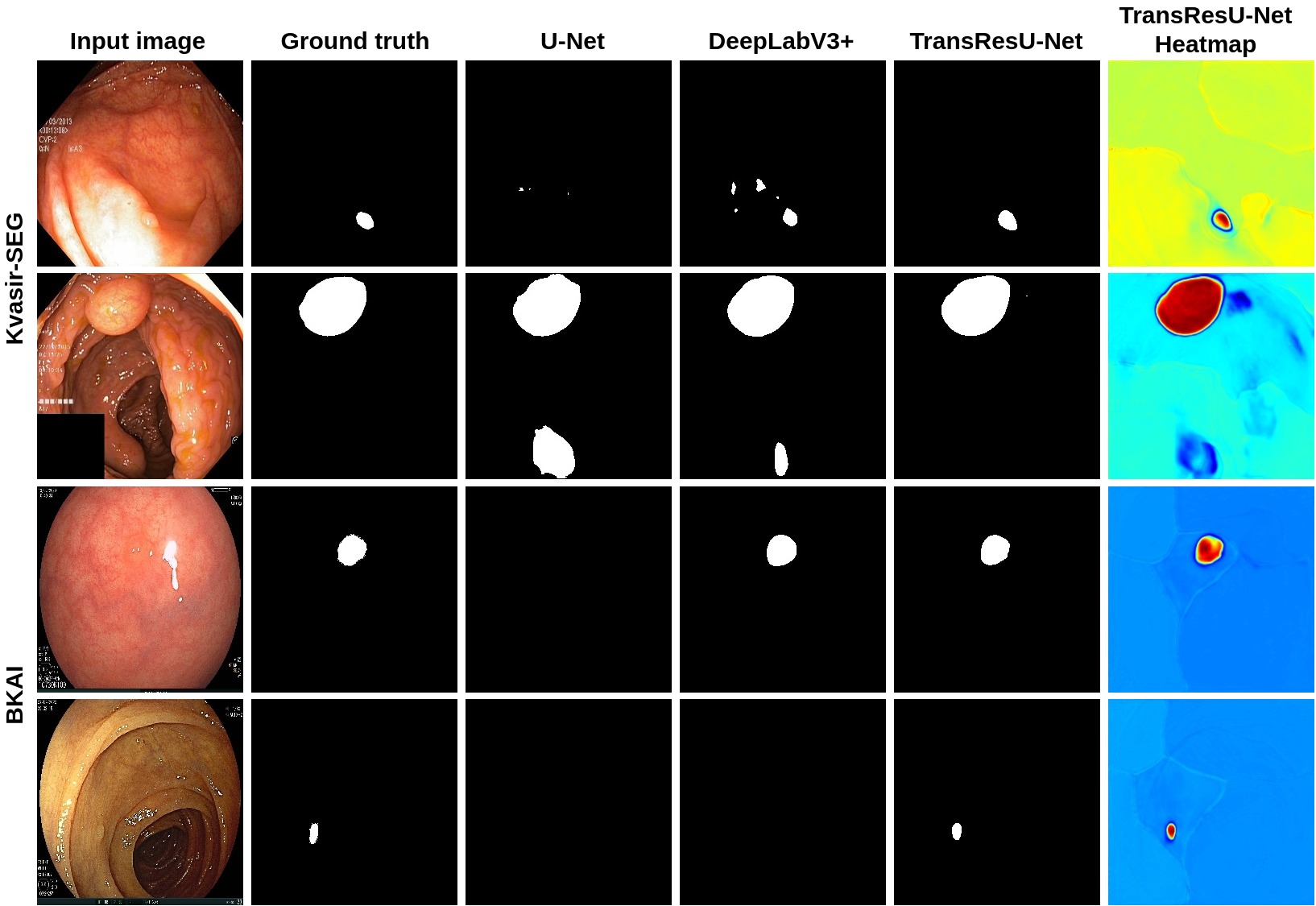}
    \caption{Qualitative results comparison along with the heatmap on the Kvasir-SEG~\cite{jha2020kvasir} and BKAI-IGH~\cite{lan2021neounet} datasets.}
    \label{fig:results}
\end{figure*}

\begin{table*} [!t]
\caption{Ablation study of the proposed TransResU-Net on the publicly available polyp datasets}
\centering
\begin{tabular}{@{}l|l|l|l|l|l|l@{}} 
\toprule
\textbf{No.} &\textbf{Dataset} &\textbf{Method} &\textbf{DSC} &\textbf{mIoU} & \textbf{Recall} & \textbf{Precision}\\ 
\midrule

\#1&\multirow{2}{7em}{Kvasir-SEG~\cite{jha2020kvasir}} &TransResU-Net (w/o Transformer Encoder block \& Dilated Conv block) &0.8679	&0.7979	&0.8863	&0.8964 \\

\#2  & &TransResU-Net (Proposed)  &\textbf{0.8884}	&\textbf{0.8214}	&\textbf{0.9106}	&\textbf{0.9022}\\

\midrule
\#1 &\multirow{2}{7em}{BKAI-IGH~\cite{lan2021neounet}} &TransResU-Net (w/o Transformer Encoder block \& Dilated Conv block) &0.8763 &0.8108 &0.8908 &0.9013 \\
\#2 & &TransResU-Net (Proposed) &\textbf{0.9154} &\textbf{0.8568} &\textbf{0.9142} &\textbf{0.9299} \\
\bottomrule
\end{tabular}
\label{table:ablation-study-table}
\end{table*}
\section{Experimental setup}
\vspace{-1mm}
We have utilized Kvasir-SEG~\cite{jha2020kvasir} and BKAI-IGH~\cite{lan2021neounet} datasets to extensively evaluate the proposed TransResU-Net. All the models used in this study are implemented using the PyTorch framework and are trained on an NVIDIA RTX 3090 GPU. The images and masks from both datasets are first resized to $256\times256$ pixels and then split into training and testing. For the Kvasir-SEG, we are using the official split, where 880 images and masks are used for training while the rest are used for testing. For the BKAI dataset, we have split the entire dataset into 80:10:10, where $80\%$ dataset is used for the training, $10\%$ is used for the validation and remaining $10\%$ is used for the testing. All the models are trained for 200 epochs with an early stopping mechanism. An Adam optimizer, learning rate of 1e$^{-4}$  with a batch size of 16 is used. A combination of binary cross-entropy loss and dice loss is used. We have trained all the models with the same set of hyperparameters for a fair comparison.

\section{Results and Discussions}
We present the quantitative results in Table~\ref{tab:results-kvasir} and Table~\ref{tab:results-bkai}.  TransResU-Net has achieved a dice coefficient of 0.8884, mIoU of  0.8214, recall of 0.9106, precision of 0.9022, accuracy of 0.9651, F2 of 0.8971 and speed of 48.61 FPS on the Kvasir-SEG. The most competitive network to TransResU-Net was DeepLabv3+~\cite{chen2018encoder} to which our architecture outperformed by 0.47\% in DSC and 0.41\% in mIoU. On the BKAI-IGH~\cite{lan2021neounet}, TransRes-UNet achieved a high DSC of 0.9154 and mIoU of 0.8568 and outperformed DeepLabv3+ by  2.17\% in DSC and 2.54\% in mIoU. 

Table~\ref{table:ablation-study-table} shows the results of ablation study. In the ablation study, we compared TransResU-Net (without Transformer encoder block and dilated convolution block) and the proposed TransResU-Net. The transformer encoder block and dilated convolution block increased the network performance by 2.05\% in DSC and 2.35\% in mIoU on the Kvasir-SEG. On the BKAI-IGH, TransUNet outperformed the prior method by 3.91\% in DSC and 4.6\% in mIoU. The recall and precision were also significantly higher for the proposed method.  Examples of qualitative results of TransResU-Net along with its heatmaps are presented in Figure~\ref{fig:results}. Here,  we show the results of UNet~\cite{ronneberger2015u}, DeepLabV3+~\cite{chen2018encoder}, and proposed TransRes-UNet on the examples such as a small or diminutive polyp, regular polyp, and flat polyp from Kvasir-SEG and BKAI-IGH dataset.  The visual comparison demonstrated that the predicted mask produced by TransResU-Net is better at delineating boundaries than DeepLabv3+ and UNet. Similarly, UNet showed under-segmentation for flat polyps whereas DeepLabv3+ showed over-segmentation for diminutive polyps. TransResU-Net could characterize all types of polyps accurately. In the qualitative results, we also show the intermediate results (heatmap) of the proposed TransResU-Net. The \textit{red} and \textit{yellow} colors in the heatmap signify the most relevant features of TransRes-UNet, whereas the \textit{blue} color shows the least significant feature produced by the architecture.  

\section{Conclusion}
In this paper, we propose the TransResU-Net architecture, which takes the advantages of transformer encoder block, residual block, and dilated convolution as its core component for real-time colonoscopy polyp segmentation. The self attention network present in the transformer, and dilated convolution block further boost the performance of the architecture. Our experimental results showed that the proposed architecture can efficiently segment polyp frames with a high dice coefficient of 0.8884 and 0.9154, respectively, on highly diverse and well-curated colonoscopy datasets. The proposed model achieved a real-time speed of 48.61 and 42.09 FPS respectively. The high performance of the algorithm on polyp segmentation tasks shows a positive signal for the development of the CADx system to be deployed in clinical settings in near future.  In the future, we plan to integrate more transformer blocks in the proposed network to further boost the performance. Additionally, we will test our algorithm on the video sequence dataset to observe if the algorithm performs reasonably well on the video sequence frames as well. 

\subsubsection*{Acknowledgement}
This project is partially supported by the NIH funding: R01-CA246704 and R01-CA240639. 

\bibliographystyle{IEEEtran}
\bibliography{references} 
\end{document}